\documentstyle[12pt]{article}
\newcommand{\be}{\begin{equation}}
\newcommand{\ee}{\end{equation}}
\newcommand{\ba}{\begin{eqnarray}}
\newcommand{\ea}{\end{eqnarray}}

\newcommand{\fr}[2]{\frac{#1}{#2}}
\newcommand{\non}{\nonumber}

\newcommand{\al}{\mbox{$\alpha$}}

\newcommand{\en}{\mbox{$\epsilon_n$}}

\newcommand{\de}{\partial}

\newcommand{\la}{\left\langle}
\newcommand{\ra}{\right\rangle}
\newcommand{\r}{\mbox{$\vec{r}$}}

\newcommand{\lv}{\mbox{$\vec{l}$}}

\newcommand{\pv}{\mbox{$\vec{p}$}}

\newcommand{\pp}{\mbox{$\vec{p}\,'$}}
\newcommand{\qv}{\mbox{$\vec{q}$}}
\newcommand{\kv}{\mbox{$\vec{k}$}}
\newcommand{\kp}{\mbox{$\vec{k}\,'$}}
\newcommand{\si}{\mbox{$\vec{\sigma}$}}

\begin{document}

\hfill BudkerINP-94-79

\hfill September 1994
\bigskip

\begin{center}{\Large \bf Order $\alpha^4 (m/M) R_{\infty}$ corrections to
hydrogen $P$ levels}\\

\bigskip

{\bf  E.A. Golosov\footnote{Novosibirsk University},
I.B. Khriplovich\footnote{e-mail address: khriplovich@inp.nsk.su},
A.I. Milstein\footnote{e-mail address: milstein@inp.nsk.su},
and A.S. Yelkhovsky\footnote{e-mail address: yelkhov@inp.nsk.su}} \\
Budker Institute of Nuclear Physics, 630090 Novosibirsk, Russia

\end{center}

\bigskip

\begin{abstract}
The order $\alpha^4 (m/M) R_{\infty}$ shift of hydrogen $P$ levels is found.
The corrections are predominantly of relativistic origin. Our approach is a
straightforward extension of that developed and applied by us previously to
positronium $P$ levels. The corrections to the Lamb shift in hydrogen
constitute numerically $\delta E(2 P_{1/2})=0.55$ kHz, $\delta E(2
P_{3/2})=0.44$ kHz.

\end{abstract}

\newpage

\section{Introduction}

Measurements of the hydrogen Lamb shift have reached now high accuracy.
Experimental values of Lamb shift for $n=2$ are

1\,057\,845(9) kHz \cite{lp},

1\,057\,851.4(1.9) kHz \cite{ps}.

The corresponding theoretical accuracy for the hydrogen Lamb shift would be
obviously very useful. In particular, the recoil corrections of the
relative order $\alpha^4 (m/M) R_{\infty}$
($R_{\infty}=109\,737.315\,682\,7(48)$ cm$^{-1}$ is the Rydberg constant)
may well turn out comparable with the quoted experimental errors. Indeed,
in positronium the order $\alpha^4 R_{\infty}$ corrections calculated in
Refs.\cite{fel,kmy} for the $2 S$ state and in Ref.\cite{kmy1} for the $2 P$
state reach 1 MHz and 0.6 MHz, respectively. The hydrogen correction addressed
in the present paper should differ from those numbers roughly by a factor
$8m/M$ where the coefficient 8 reflects the dependence of the positronium
result on the reduced mass $m/2$ which enters the shift at least in the third
power. In this way we come to the conclusion that the discussed
corrections in hydrogen can well constitute few kHz.

The $\alpha^4 (m/M) R_{\infty}$ correction to the hydrogen $2S$ states has
been found recently\cite{gp}, and constitutes -0.92 kHz. As to hydrogen
$P$ states, the calculation of their shift can be done easily within the
approach developed and applied by us earlier to positronium $P$ states
\cite{kmy1}. This is the subject of the present paper.

Recoil corrections emerge from two sources. Some effective operators
contain $M^{-1}$ explicitly. When treating other perturbations, independent
of $M$, order $m/M$ corrections originate from the dependence on the
reduced mass $\mu$ of nonrelativistic wave functions, entering the
expectation values.

Major part of corrections is of relativistic nature. As for the true radiative
corrections of the order discussed, for the states of nonvanishing orbital
angular momentum they originate from the electron anomalous magnetic moment
only, as it was assumed in Refs.\cite{dk,kmy} and proven accurately in
Ref.\cite{ekm}.

\section{Contributions of irreducible operators}

\subsection{Relativistic correction to the dispersion law}

Let us start with the kinematic correction, generated by the $v^4/c^4$ term
in the dispersion law for the electron,
\ba
\sqrt{m^2 + p^2} - m &=& \fr{p^2}{2m} - \fr{p^4}{8m^3} + \fr{p^6}{16m^5} +
                         \ldots , \\
V^{(1)}_{kin} &=&  \fr{p^6}{16m^5} .
\ea
To calculate the corresponding expectation value is a simple problem in
quantum mechanics. So,
\ba
E^{(1)}_{kin} &=& -\fr{m^2}{M}\fr{\de}{\de\mu}\la \fr{p^6}{16m^5} \ra \\
&=&- \fr{\en}{5}\left( 8 - \fr{17}{n^2} + \fr{75}{8n^3} \right).
\ea
Here
$$\mu=\frac{mM}{M+m} \approx m(1-\frac{m}{M})$$
is the hydrogen reduced mass;
$$\en \equiv \frac{m^2 \al^6}{M n^3}.$$
The result differs from that of Ref.\cite{kmy1} for positronium by a fairly
obvious scaling factor.

\subsection{Relativistic corrections to the Coulomb interaction}

This perturbation operator, as extracted from the $(v/c)^4$ corrections to
the Coulomb scattering amplitude for free particles, equals
\be\label{eq:C+}
V_{C} = - \fr{\al}{32m^4} \fr{4\pi}{q^2} \left\{ \frac{5}{4} (p'^2-p^2)^2 -
 3i(\vec{\sigma},\pp\times\pv) (p^{\prime 2}+p^2) \right\}.
\ee
We are neglecting systematically here and below effective operators
proportional to $\delta(\r)$ in the coordinate representation, their
expectation values vanishing for $P$ states. This energy correction is
\ba
E^{(1)}_{C} &=& \en \left\{\frac{5}{16} \left( 1 -
\fr{2}{3n^2} \right)
+ \frac{3}{4} (\vec{\sigma} \vec l)\left( 1 - \fr{13}{12n^2} \right) \right\}.
\ea
Calculational details pertinent to the problem can be found in Ref.\cite{kmy1}.

Now, due to the Coulomb interaction electron can go over into
a negative-energy intermediate state. The corresponding contributions are
described by $Z$-diagrams of the kind presented in Fig.1.
The corresponding perturbation operator is
\be
V_{C-} = - \fr{(4\pi\al)^2}{8m^3} \int \fr{d^3 k}{(2\pi)^3}\;
              \fr{\vec{k}(\qv-\vec{k})}{k^2(\qv - \vec{k})^2}.\label{eq:C-}
\ee
The energy correction generated in this way equals
\be
E^{(1)}_{C-} = - \fr{\en}{5} \left( 1 - \fr{2}{3n^2} \right).
\ee

\subsection{Single magnetic exchange}

In the noncovariant perturbation theory the electron-proton scattering
amplitude due to the exchange by one magnetic quantum is
\be\label{eq:AM}
A_M = - \fr{4\pi\al}{2q} j_i (\pp, \pv) J_j (-\pp, -\pv)
\left (\frac{1}{E_n-q-p^2/2m} + \frac{1}{E_n-q-p'^2/2m}\right)
\left( \delta_{ij} - \fr{q_i q_j}{q^2} \right).
\ee
In the dispersion law for electron it is sufficient here to confine to the
nonrelativistic approximation. The proton current to our accuracy reduces to
\be
\vec{J} (-\pp, -\pv\,)=-\frac{1}{2M}(\pp+\pv\,)
\ee
We omit at the moment the hyperfine contributions induced by
the spin part of this current
\be
\vec{J}^s (-\pp, -\pv)=-\frac{1}{2M}\; ig[\vec{\sigma}_p \times
(\pp-\pv)]
\ee
where $g=2.79$ is the proton magnetic moment. All nuclear-spin-dependent
effects will be discussed below.

In the electron current we have to keep the $(v/c)^2$ corrections:
\be\label{eq:j}
\vec{j}(\pp, \pv \,) = \frac{1}{2m}\{\pp+\pv+i[\vec{\sigma}\times
(\pp-\pv)]\}\left(1 - \frac{p^{\prime 2} + p^2}{4m^2}\right)
- \frac{(p^{\prime 2} - p^2)^2}{16m^3} i[\vec{\sigma}\times
(\pp+\pv\,)].
\ee
They produce the following energy shift:
\ba
E^{(1)}_{curr} &=& \la\fr{\al}{4mM} \fr{4\pi}{q^2} \left\{
             \fr{(p'^2-p^2)^2}{4m^2}
             \fr{i(\si,\qv\times\pv)}{q^2} \right. \right.\non \\
             && \left.\left. + \fr{p'^2+p^2}{2m^2} \left(
             2\fr{(\qv\times\pv)^2}{q^2}
             +i(\si,\qv\times\pv) \right) \right\}\ra \\
             &=& \en \left\{ \fr{7}{15} - \fr{31}{30n^2} + \fr{1}{2n^3}
                 - \fr{\si\lv}{4} \left( 1 - \fr{1}{n^2} \right)\right\}. \non
\ea

Let us consider now the retardation effect. To this end the currents can be
taken in the leading approximation, while the perturbation of interest
originates from the second-order term of the expansion of the factor
$[E_n-p^2/2m-q]^{-1}$ in (\ref{eq:AM}) in powers of $(E_n-p^2/2m)/q$,
\ba
E^{(1)}_{ret} &=& \la- \fr{\al}{4mM} \fr{4\pi}{q^2} \fr{\left( E_n -
            p^2/2m \right)^2 + \left( E_n - p^{\prime 2}/2m
            \right)^2}{q^2} \right.\\
        &&\cdot \left.\left\{ 2\fr{(\qv\times\pv)^2}{q^2} + i(\si,\qv\times\pv)
 \right\} \ra \non \\
          &=& \en \left\{ \fr{2}{5} - \fr{1}{4n} + \fr{3}{20n^2}
                 + \fr{\si\lv}{30} \left( 4 - \fr{1}{n^2} \right) \right\}.
\ea

Magnetic quantum propagates for a finite time and can cross arbitrary number
of the Coulomb ones. Simple counting of the momenta powers demonstrates
that it is sufficient to include the diagrams with one and two Coulomb
quanta (dashed lines) crossed by the magnetic photon (wavy line). In the
first case, Fig.2, the perturbation operator arises as a product of the
Pauli currents and the first-order term in the expansion in
$(E_n-p^2/2m)/q$:
\ba\label{eq:MC}
E^{(1)}_{MC} &=& \la -(4\pi\al)^2 \int \fr{d^3 k}{(2\pi)^3}
  \fr{\delta_{ij}-\fr{k_ik_j}{k^2}}{2k(\qv- \kv)^2} \non \right.\\
          && \left( J_i(\pv,\pv+\kv) \,j_j(\pp,\pp+\kv)\;
           \fr{2E_n-(\vec{p}\,^{\prime}-\vec{k})^2/2m-p^2/2m}{k^3}
           \non \right.\\
&& \left.\left.+ J_i(\pp,\pp+\kv)\,j_j(\pv,\pv+\kv)\;
           \fr{2E_n-(\vec{p}+\vec{k})^2/2m-p^{\prime 2}/2m}{k^3}
           \right)\ra \\
         &=& \en \left\{ -\fr{13}{20} + \fr{1}{2n} - \fr{3}{20n^2}
         - \si\lv\left(\frac{7}{60} + \frac{1}{30n^2}\right)\right\}
\ea
In the second case all the elements of diagram 3 should be taken
to leading nonrelativistic approximation:
\ba\label{eq:MCC}
E^{(1)}_{MCC} &=& \la - (4\pi\al)^3 \int \fr{d^3 k}{(2\pi)^3}
         \int \fr{d^3 k'}{(2\pi)^3}\;\fr{\delta_{ij}-k_ik_j/k^2}{2k^4
         (\qv-\kp)^2 (\kp-\kv)^2 } \right. \non \\
        && \left.\left\{ J_i(\pv,\pv+\kv)\, j_j(\pp,\pp+\kv)
           +J_i(\pp,\pp+\kv)\,j_j(\pv,\pv+\kv)
           \right\}\ra \\
       &=& \fr{\en}{4}\left\{\fr{5}{3}-\fr{1}{n}+\fr{\si\lv}{3}\right\}.
\ea

One more energy correction of the $\al^4 R_{\infty}m/M$ order at the single
magnetic exchange is due to the electron transitions to negative-energy
intermediate states (see Fig.4). To leading approximation one gets easily
\ba
E^{(1)}_{M-} &=& \la \fr{\al^2}{4mM} \int \fr{d^3 k}{(2\pi)^3}
           \;\fr{(4\pi)^2}{k^2 (\qv - \kv)^2}\; i(\si,\kv\times\pv)\ra\\
              &=&-\,\fr{\en}{10}\left( 1 - \fr{2}{3n^2} \right)\si\lv.
\ea

\subsection{Double magnetic exchange}

Let us consider now irreducible diagrams with two magnetic quanta. To our
approximation they are confined to the type presented in Fig.5. Their sum
reduces to
\ba\label{eq:MM}
E^{(1)}_{MM} &=& \la \fr{\al^2}{2m^2M} \int \fr{d^3 k}{(2\pi)^3}
           \;\fr{(4\pi)^2}{k^2 k'^2}
           \left\{ \pv\pp - 2\fr{(\kv\pv)(\kv\pp)}{k^2} +
           \fr{(\kv\pv)(\kv\kp)(\kv\pp)}{k^2 k'^2} - \fr{\kv\kp}{2}  \right.
           \right.\non \\
       &&  \left.\left. + i\si \left( \kp\times\pv -
           \fr{\kp\times\kv\;(\kv\pv)}{k^2} \right) \right\} \ra \\
         &=& \en \left\{\fr{1}{3}\left(1-\fr{1}{n^2}\right)
         - \fr{\si\lv}{10}
         \left(1- \fr{2}{3n^2}\right)\right\},
\ea
Here $\kp = \qv - \kv$.

\section{Corrections of second order \newline in the Breit Hamiltonian}

Next class of the order $\alpha^4 R_{\infty}$ corrections originates from
the iteration of the usual Breit Hamiltonian $V$ of second order in
$v/c$.

Omitting again nuclear-spin-dependent terms and those with $\delta (\vec r)$,
we present the Breit perturbation for hydrogen
(see, e.g., \cite{blp}, \S 84),
\be \label{eq:VB}
V = - \fr{p^{4}}{8m^{3}}
      +\fr{\al}{4m^{2}r^{3}}\,\si\lv
    - \fr{\al}{2mMr}\left(p^{2}+\fr{1}{r^2}\r(\r\pv)\pv\right)
 + \fr{\al}{2mMr^3}\,\si\lv
\ee
as:
\ba\label{rp}
V & = & m\al^4 v, \\
v & = & a\left\{h,\frac{1}{r}\right\} + b\, [ h, ip_r]
      + c\, \frac{1}{r^2}. \label{eq:v}
\ea
Here
\be\label{ko}
a = -\fr{1}{2} + \fr{m}{M}, \;\; b = \fr{\si\lv}{8} +
\fr{m}{M}\left(\fr{1}{2}
-\fr{\si\lv}{8}\right), \;\; c = -\fr{1}{2} + b;
\ee
$p_r = -i ( \de_r + 1/r )$ is the radial momentum, while
\[ h = \fr{p_r^2}{2} + \fr{1}{r^2} - \fr{1}{r} \]
is the unperturbed hydrogen Hamiltonian for the radial motion with $L=1$,
written in atomic units.

It is a simple quantum-mechanical exercise now to derive the second-order
energy correction from perturbation (\ref{rp}). The details of derivation,
as applied to positronium, are presented in Ref.\cite{kmy1}. In the
hydrogen case the result reads
\ba
\Delta E &=& \fr{m^2 \alpha^6}{4\mu n^3} \left\{ -\fr{3a^2+14ab+13b^2}{15} -
        \fr{2c(2c+9a+9b)}{27} \right. \nonumber \\
   && \left. - \fr{2c^2}{3n} + \fr{2}{3n^2}\left( \fr{11a^2 + 13ab +
      6b^2}{5} + 4ac \right) - \fr{5a^2}{2n^3} \right\}.
\ea
We substitute now into this expression values (\ref{ko}) for
$a,b,c$ and single out terms $\sim M^{-1}$ of interest to us. The
result is

\be
E^{(2)} = \en \left\{ \fr{467}{480} + \fr{3}{16n} - \fr{347}{120n^2}
 + \fr{15}{8n^3} - \si\lv \left(\fr{419}{960} + \fr{3}{32n}
 -\fr{53}{80n^2} \right)\right\}.
\ee

\section{Corrections to the hyperfine interaction}

Let us discuss now the energy corrections induced by the magnetic
interaction with the proton magnetic moment, i.e. relativistic corrections
to the hyperfine interaction. The complete relativistic expression for the
hyperfine level splitting in hydrogen reduces to the expectation value of
the interaction between the relativistic electron and nuclear magnetic
moment calculated with the Dirac wave functions:
\be\label{dir}
\delta E_{hfs}(nlj;F)= \alpha \vec{\mu}\,\la nlj\left|\frac{\vec r}{r^3}
\times\vec{\alpha}\right|nlj \ra
\ee
(here $\vec{\mu}$ is the nuclear magnetic moment operator, which is by itself
$\sim 1/M$, $\vec{\alpha}$ are the Dirac $\alpha$-matrices. This formula can be
derived\cite{sh}, practically in the same way as the Dirac equation itself,
from the analysis of the Feynman diagrams.

The non-trivial, radial part of this expectation value can be conveniently
calculated by means of a virial relation (see, e.g., Ref.\cite{sh1}), and
the final result reads
\be\label{dir1}
\delta E_{hfs}(nlj;F)= \frac{\vec{\mu}\cdot\vec j}{j(j+1)}\, [j(j+1) -l(l+1)
+ 1/4] \,\alpha^2\, \frac{\partial E_{nj}}{\partial \kappa}\; \frac{E_{nj} -
m/2\kappa}{j(j+1) - \alpha^2}
\ee
Here $E_{nl}$ is the eigenvalue of the Dirac Coulomb problem, $\kappa =
(l-j)(2j+1)$.

Of course, the discussed relativistic correction to the hyperfine
interaction can be derived also in the same way as that independent of
nuclear spin (see the previous sections). In such an approach the
contributions due to the retardation of the magnetic interaction and to
diagrams 2 and 3 cancel out, which just corresponds to the instantaneous
nature of the magnetic interaction implied by formula (\ref{dir}). The
final result of this calculation (it is presented in the last section)
coincides of course with the $\alpha^2$-expansion of formula (\ref{dir1}).

\section{True radiative corrections}

Even the true radiative corrections of the $\al^4 R_{\infty}m/M$
order to $P$-levels can be presented in a simple form
practically without special calculation. It was suggested in
Refs.\cite{dk,kmy1} and accurately proven in Ref.\cite{ekm} that all true
radiative corrections to levels of $l\neq 0$ are confined to the electron
anomalous magnetic moment contributions to the single magnetic exchange
and to the spin-orbit interaction\footnote{Unfortunately, in the
paper\cite{kmy1} on positronium by three of us, the contribution of the
electron anomalous magnetic moment to the spin-orbit interaction was lost.
This correction to positronium $P$-levels equals
$$-\,\frac{0.328}{\pi^2}\,\frac{m\alpha^6}{24n^3}\,\vec L\vec S.$$ It
constitutes 0.0032, 0.0016 and --0.0016 MHz at $j$=0, 1 and 2
respectively, which is too small to influence the overall numerical
results.}.

It can be easily demonstrated that only the second-order correction to the
electron anomalous magnetic moment, $-0.328\alpha^2/\pi^2$, contributes to
the order $\alpha^4 (m/M) R_{\infty}$ shifts of hydrogen levels
with $l\neq 0$. In particular, as well as in positronium, the anomalous
magnetic moment contributions to the first-order retardation effect and to
diagram 2 cancel.

In this way we come to the following expression for the radiative shift of
hydrogen $nP$ levels
\be
E^{(1)}_{rad} = \en\, \fr{0.328}{\pi^2}
                \left\{ \frac{1}{3}\,\si\lv
    + 2.79\, \fr{(\si\lv)(\vec j \vec{\sigma}_p)}{12 j(j+1)}\right\}.
\ee

\section{Results}

The total correction to hydrogen $P$-levels independent of nuclear spin is
\be
\delta E(nP_j) = \en \left\{ \fr{217}{480} + \fr{3}{16n} - \fr{14}{15n^2}
 + \fr{1}{2n^3} - \si\lv \left(\fr{7}{192} + \fr{3}{32n}
 -\fr{1}{6n^2} - \fr{1}{3} \fr{0.328}{\pi^2} \right)\right\}.
\ee
Its numerical values at $n=2$ constitute
$$\delta E(2P_{1/2}) = 0.55 kHz,$$
$$\delta E(2P_{3/2}) = 0.44 kHz.$$
They are somewhat smaller than our crude estimates outlined in Introduction.

The hyperfine corrections at given total atomic angular momentum $F$ can
be presented in an analogous form:
\be
\delta E(nP_j;F) = \en\, 2.79\, \fr{\vec j \vec{\sigma}_p}{2j(j+1)}
\left\{ \fr{157}{270} + \fr{2}{3n} - \fr{7}{5n^2}
- \si\lv \left(\fr{173}{540} + \fr{1}{6n}
 -\fr{2}{15n^2} - \fr{1}{6}\, \fr{0.328}{\pi^2} \right)\right\}.
\ee
Numerically these contributions to the hyperfine splitting of $2P_j$-levels
constitute
$$\Delta_{hf}(2P_{1/2}) = 6.12 kHz,$$
$$\Delta_{hf}(2P_{3/2}) = 0.38 kHz.$$

\bigskip
{\bf Acknowledgements}

We are grateful to H. Grotch for the communication of the results of
Ref.\cite{gp} prior to publication. We acknowledge the support by the
Program "Universities of Russia", Grant No. 94-6.7-2053.

\newpage

{\bf Figure captions}

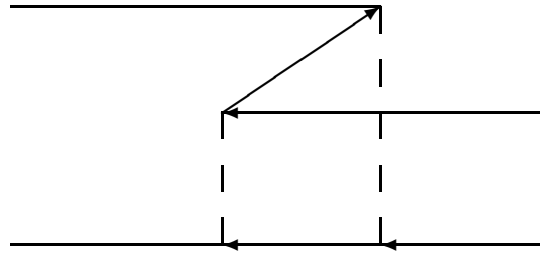
\noindent {\it Fig.1.} Z-type double-Coulomb exchange

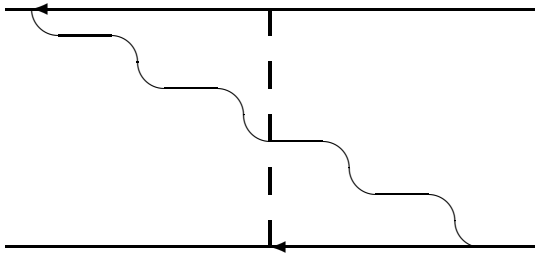
\noindent {\it Fig.2.} Single-magnetic-single-Coulomb exchange

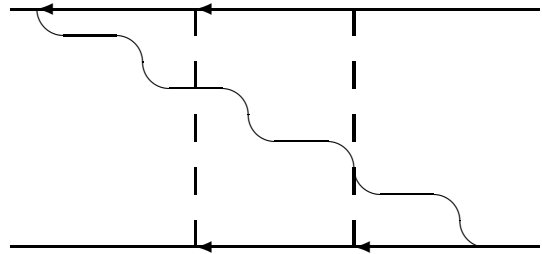
\noindent {\it Fig.3.} Single-magnetic-double-Coulomb exchange

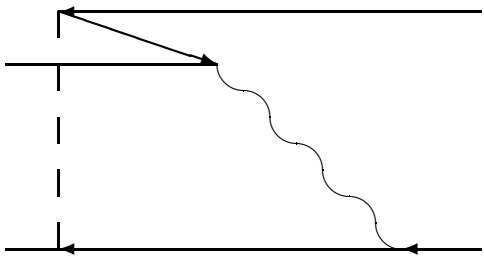
\noindent {\it Fig.4.} Z-type single-magnetic-single-Coulomb exchange

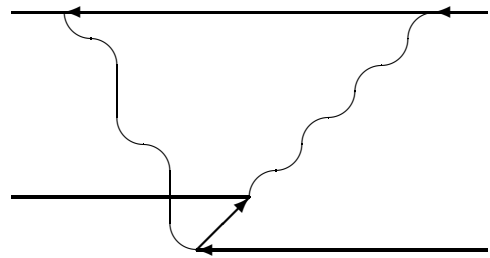
\noindent {\it Fig.5.} Double-magnetic exchange

\newpage

\begin{figure}
 \begin{picture}(460,180)
  \put(115,40){\begin{picture}(230,140)
              \thicklines
              \put(10,110){\line(1,0){140}}
              \put(210,70){\vector(-1,0){120}}
              \put(90,70){\vector(3,2){60}}
              \put(210,20){\vector(-1,0){60}}
              \put(150,20){\vector(-1,0){60}}
              \put(90,20){\line(-1,0){80}}
              \multiput(90,20)(0,20){3}{\line(0,1){10}}
              \multiput(150,20)(0,20){5}{\line(0,1){10}}
              \put(90,0){Fig.1}
            \end{picture}}
 \end{picture}
\end{figure}

\begin{figure}
 \begin{picture}(460,180)
  \put(0,40){\begin{picture}(230,140)
             \multiput(200,30)(-40,20){5}{\oval(40,20)[bl]}
             \multiput(160,30)(-40,20){4}{\oval(40,20)[tr]}
             \thicklines
             \multiput(210,20)(-90,90){2}{\vector(-1,0){100}}
             \multiput(210,20)(0,90){2}{\line(-1,0){200}}
             \multiput(110,20)(0,20){5}{\line(0,1){10}}
             \put(90,0){Fig.2}
             \end{picture}}
  \put(230,40){\begin{picture}(260,140)
              \multiput(200,30)(-40,20){5}{\oval(40,20)[bl]}
              \multiput(160,30)(-40,20){4}{\oval(40,20)[tr]}
              \thicklines
              \multiput(200,20)(-60,90){2}{\vector(-1,0){60}}
              \multiput(140,20)(-60,90){2}{\vector(-1,0){60}}
              \multiput(10,20)(0,90){2}{\line(1,0){200}}
              \multiput(80,20)(0,20){5}{\line(0,1){10}}
              \multiput(140,20)(0,20){5}{\line(0,1){10}}
              \put(90,0){Fig.3}
              \end{picture}}
 \end{picture}
\end{figure}

\begin{figure}
 \begin{picture}(460,180)
  \put(0,40){\begin{picture}(230,140)
            \multiput(160,30)(-20,20){4}{\oval(20,20)[bl]}
            \multiput(140,30)(-20,20){3}{\oval(20,20)[tr]}
            \thicklines
            \multiput(190,20)(0,90){2}{\vector(-1,0){160}}
            \put(190,20){\vector(-1,0){30}}
            \put(30,110){\vector(3,-1){60}}
            \multiput(90,20)(0,70){2}{\line(-1,0){80}}
            \multiput(30,20)(0,20){5}{\line(0,1){10}}
            \put(90,0){Fig.4}
            \end{picture}}
  \put(230,40){\begin{picture}(260,140)
               \multiput(190,110)(-20,-20){4}{\oval(20,20)[tl]}
               \multiput(170,110)(-20,-20){3}{\oval(20,20)[br]}
               \multiput(60,80)(20,-40){2}{\oval(20,60)[tr]}
               \multiput(60,120)(20,-40){3}{\oval(20,20)[bl]}
               \thicklines
               \multiput(160,120)(50,-90){2}{\vector(-1,0){110}}
               \put(210,120){\vector(-1,0){20}}
               \put(100,30){\vector(1,1){20}}
               \multiput(70,120)(120,0){2}{\line(-1,0){40}}
               \put(30,50){\line(1,0){90}}
               \put(90,0){Fig.5}
               \end{picture}}
 \end{picture}
\end{figure}

\end{document}